\documentclass[11pt,draftclsnofoot,onecolumn]{IEEEtran}

\usepackage[english]{babel}

\usepackage[letterpaper,top=2cm,bottom=2cm,left=3cm,right=3cm,marginparwidth=1.75cm]{geometry}

\usepackage{amsmath}
\usepackage{graphicx}
\usepackage[colorlinks=true, allcolors=blue]{hyperref}

\usepackage{subcaption}
\usepackage{overpic}
\usepackage{cite}

\title{25 Years of Signal Processing Advances for Multiantenna Communications}

\author{\IEEEauthorblockN{Emil Bj\"ornson, Yonina C. Eldar, Erik G. Larsson, Angel Lozano, H. Vincent Poor}}

\begin{document}

\maketitle

Wireless communication technology has progressed dramatically over the past 25 years, in terms of societal adoption as well as technical sophistication. 
In 1998, mobile phones were still in the process of becoming compact and affordable devices that could be widely utilized in both developed and developing countries. There were ``only'' 300 million mobile subscribers in the world \cite{ITU1999}.
Cellular networks were among the first privatized telecommunication markets, and competition turned the devices into fashion accessories with attractive designs that could be individualized.
The service was circumscribed to telephony and text messaging, but it was groundbreaking in that, for the first time, telecommunication was between people rather than locations.

There are now more than six billion subscribers worldwide, and the mobile phone remains the main wireless device, but much has changed.
Traditional feature phones with physical keypads have been replaced by smartphones with large touchscreens.
Telephony nowadays constitutes a negligible fraction of the traffic, the vast majority of which amounts to packets bearing data for end-user applications. Video and audio streaming, social media, gaming, and a host of other apps, generate the bulk of the traffic. New services continue to arise and cement the smartphone's central role in nearly every aspect of our lives. In parallel, non-human-operated devices are progressively coming online to form the Internet-of-things (IoT) as society continues to be digitized.

Wireless networks have changed dramatically over the past few decades, enabling this revolution in service provisioning and making it possible to accommodate the ensuing dramatic growth in traffic.
There are many contributing components, including new air interfaces for faster transmission, channel coding for enhanced reliability, improved source compression to remove redundancies, and leaner protocols to reduce overheads.
Signal processing is at the core of these improvements, but nowhere has it played a bigger role than in the development of multiantenna communication.
This article tells the story of how major signal processing advances have transformed the early multiantenna concepts into mainstream technology over the past 25 years. The story therefore begins somewhat arbitrarily in 1998.
A broad account of the state-of-the-art signal processing techniques for wireless systems by 1998 can be found in \cite{poor1998wireless}, and its contrast with recent textbooks such as \cite{Marzetta2016a,massivemimobook,heath2018foundations} reveals the dramatic leap forward that has taken place in the interim.

\section*{Fundamentals of Multiantenna Communications}

\begin{figure} 
        \centering 
        \begin{subfigure}[b]{0.5\columnwidth} \centering
	\begin{overpic}[width=\columnwidth,tics=10]{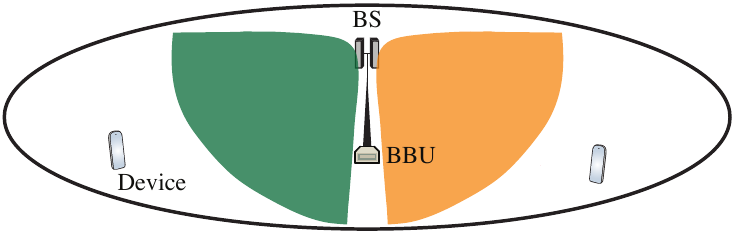}
\end{overpic} 
                \caption{Typical 2G deployment.} 
        \end{subfigure}%
        \begin{subfigure}[b]{0.5\columnwidth} \centering 
	\begin{overpic}[width=\columnwidth,tics=10]{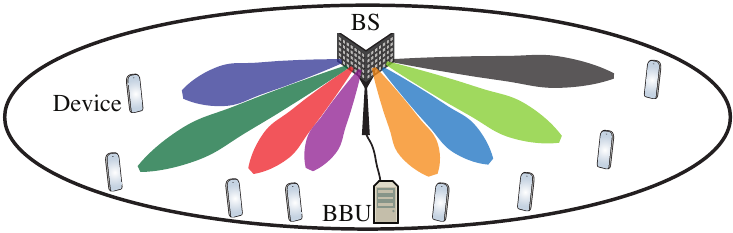}
\end{overpic} 
                \caption{Typical 5G deployment.} 
        \end{subfigure} 
        \caption{(a) 2G deployment in 1998, consisting of fixed directive antennas that broadcast each signal into a sector. (b) 5G deployment in 2023, entailing antenna arrays that can exploit the three main multiantenna benefits:
        beamforming gain, spatial diversity, and spatial 
        multiplexing.} 
        \label{fig:MIMO-fundamentals}  
\end{figure}

Traditionally, a base station (BS) at a cellular network site featured antenna panels connected to a baseband unit (BBU) that managed the digital signal processing. These panels, in turn, were tall and narrow, containing multiple vertically stacked radiating elements.
By emitting the same signal from such elements, constructive superposition was leveraged to create a radiation pattern, vertically narrow and horizontally wide, that covered a swath of ground in a predefined manner.
This is illustrated in Fig.~\ref{fig:MIMO-fundamentals}(a), with each panel's coverage region termed a cell sector.

At current BSs, the panels have been replaced with antenna arrays having a more symmetric aspect ratio,
which results in radiated beams that can be narrow both horizontally and vertically. The signal transmitted from each antenna element is individually controlled by the BBU, which now has far stronger computational capabilities and can alter the physical shape of the produced beam over both time and frequency. 
Fig.~\ref{fig:MIMO-fundamentals}(b) illustrates such a setup, and how each beam is narrow enough to aim at a particular user.
When these arrays are used in propagation environments with 
multiple widely spaced paths, each radiated signal loses its directional beam-shape and is instead fine-tuned to make the paths superimpose coherently on a small region around the intended receiver.

Antenna arrays bring about three main categories of benefits:
\begin{enumerate}
    \item \textbf{Beamforming gain:} The transmit beam is focused on the receiver, whereby a larger fraction of the radiated energy reaches it. Likewise, multiple receive antennas can collect more energy from selected directions, reinforcing the beam at that end with a focus on the transmitter. The overall beamforming gain is proportional to the transmit and receive array sizes.
    
    \item \textbf{Spatial diversity:} There are generally multiple paths via which signals travel between the transmitter and receiver, and the ensuing signal replicas can combine destructively. This causes signal fading, which antenna arrays can mitigate by observing multiple fading realizations simultaneously.
    
    \item \textbf{Spatial multiplexing:} Multiple signals can be transmitted concurrently on different beams, either to a single user equipped with multiple antennas, or to multiple users as in Fig.~\ref{fig:MIMO-fundamentals}(b). This provides a traffic multiplier or \emph{multiplexing gain}, provided the interference among the signals can be kept at bay. 
\end{enumerate}

Above, and in the sequel, the beamforming gain is taken as the increase in signal power at the receiver, yet a more nuanced description would further include the reduction in interference to and from unintended users \cite[Sec.~5.7]{heath2018foundations}. With a careful design, beamforming can strike an optimum balance between increasing signal energy and reducing interference.

\section*{State-of-the-Art in 1998}

Some of the benefits of antenna arrays were understood well before 1998, but their technology readiness levels were much different than today.
Marconi himself famously capitalized on beamforming to enable wireless transatlantic communication in 1901.
That experiment relied on an \emph{array antenna}, which achieves beam directivity by connecting multiple elements to the same signal generator. The geometry of the array antenna determines the direction in which the radiated signals superimpose constructively. Hence, the beam direction is fixed and determined at the time of building and erecting the array.
This is how the 2G antennas in Fig.~\ref{fig:MIMO-fundamentals}(a) were designed to cover a sector with a fixed beam.

\begin{figure}[t!]
        \centering 
	\begin{overpic}[width=.95\columnwidth,tics=10]{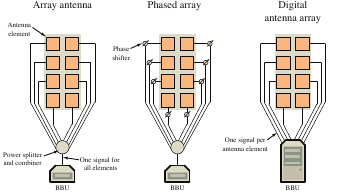}
\end{overpic} 
        \caption{There are three classical categories of arrays: 1) \emph{array antennas} that generate fixed beams; 2) \emph{phased arrays} that rely on phase shifters to control the beam direction; 3) \emph{digital antenna arrays} that have full control of the signal transmitted from each antenna element.} 
        \label{fig:array_antennas}  
\end{figure}

A different beam direction than the one dictated by the array geometry can be realized by emitting the same signal from all the elements, but with appropriate phase shifts. This concept was first observed experimentally by Ferdinand Braun in 1902, and it led to the \emph{phased array} technology used for radar since World War II. The phase shifts can be varied over time, to scan for objects in different angular directions.
Early field trials of phased arrays for 2G were conducted in 1996 \cite{Anderson1996a}.
The possibility of pointing the beams to user locations
opened the door to stronger directivities and higher gains, since a beam no longer had to cover an entire sector.
The difference between array antennas and phased arrays is illustrated in Fig.~\ref{fig:array_antennas}, which also depicts the \emph{digital antenna arrays} featured in 5G, where each element is connected to a separate signal generator.

In parallel with the refinement of phased arrays for beamforming over several decades, the use of multiple receive antennas for diversity became also commonplace.
Spatial diversity was conceived for signal reception as far back as the 1930s \cite{Peterson1931a} and builds on an intuitive principle: if the same signal reaches several physically separated antennas, it is unlikely that the multipath propagation environment causes destructive superposition, hence signal fading, at all such antennas simultaneously. By decoding a combination of the observations at the
various receive antennas, the communication becomes much more reliable.

A wireless network must of course provide an uplink connection from users to BSs, as well as a  downlink connection from BSs to users. Thus, diversity is desirable in both link directions, yet transmit diversity did not emerge until the 1990s \cite{wittneben1991basestation}.
In 1998, the Alamouti space-time block code for two antennas was proposed \cite{Alamouti1998a} and a more general framework for space-time coding with multiple transmit antennas was published soon thereafter \cite{Tarokh1998a}.
The principle is to repeatedly transmit a block of data symbols while varying the spatial directivity in a predetermined way (e.g., using different antennas); the receiver collects observations over a time interval and decodes them.
Space-time codes are carefully crafted to not only enable decoding, but to strike a satisfactory tradeoff between high spectral efficiency (i.e., bits per second per Hertz of spectrum), high diversity, and low complexity.

Altogether, beamforming gains and spatial diversity were known by 1998, and it was largely thought that these were the two main benefits of antenna arrays: beamforming gains in the case of coherent arrays, associated with tight antenna spacings and cleanly defined directions of arrival and departure, and diversity gains in the case of arrays experiencing largely uncorrelated fading across the antennas, associated with wider spacings and rich multipath settings.
The third, and ultimately the most powerful benefit of antenna arrays, spatial multiplexing, was still largely under the radar. However, its seeds had already been planted in research efforts on interference-aware beamforming \cite{Winters1984a} and on communication concepts for linear channels that couple multiple inputs into multiple outputs \cite{Sal:Digital-transmission-over:85}.
Unlike beamforming and diversity, which involved replicas of a single signal, these precursors of spatial multiplexing entailed the transmission and reception of distinct signals simultaneously and on the same bandwidth.
Particularly prescient was the transmission and reception with two orthogonally polarized antennas, subsequently extended in a piece that featured multiantenna transmitters and receivers with many of the ingredients required for true spatial multiplexing \cite{Win:On-the-Capacity-of-Radio-Communication:87}. However, it was not until after 1998 that all these pieces fell into place.

\subsection*{External Technology Developments}

Three external trends have heavily guided and influenced the evolution of multiantenna technology over the past decades.

\begin{itemize}
\item The explosion in wireless traffic, which has doubled every 18 months as per Cooper's law, along with a fundamental change in the nature of such traffic, driven by new user behaviors and applications.
An efficient network for telephony had to support many simultaneous fixed-rate connections, while today's data networks aim at maximizing the bit rate per user device (to support certain applications) and the bit rate per unit area (to accommodate many devices).

\item The exponential improvement and size reduction of integrated circuits have led to systems-on-a-chip that combine radios, memory, and processors capable of advanced signal processing on a tiny piece of silicon.
While, in 1998, a digital antenna array with $M =2$ or $M=4$ elements would consist of $M$ external antenna elements connected to $M$ radio frequency (RF) units and one BBU, current 5G BSs can integrate $M=64$ elements and RF units into a single box.
This development has also enabled smartphones to feature digital arrays, for now with $M=4$ elements.

\item The gradual change in the signal waveforms. There were multiple 2G standards based either on time-division multiple access (TDMA) or code-division multiple access (CDMA).
The first versions of 3G, finalized precisely around 1998, were entirely based on CDMA, which won the battle against the competing orthogonal frequency-division multiple access (OFDMA).
For 4G, the shift to OFDMA finally took place, and 5G retained this same waveform after a handful of alternatives were evaluated and discarded. 
While all waveforms are in principle compatible with antenna arrays, the choice does have a fundamental impact on what signal processing algorithms are required.
\end{itemize}

\section*{Five Key Areas of Signal Processing Advances}

We have identified five stages of signal processing advances in the evolution of multiantenna technology from 2G to 5G and beyond.
The background, new solutions, and specific insights are expounded on in the following sections.

\subsection*{From Spatial Diversity to Spatial Multiplexing}

One could argue that, all the way back to Marconi, beamforming was motivated by the interest in extending the range of coverage. In turn, diversity was motivated by the desire to increase reliability. By 1998, the exploding cost of radio spectrum ahead of 3G brought about a new and powerful necessity: increasing the spectral efficiency. The shift towards high-bit-rate user applications further amplified this trend.
The operational mode of antenna arrays that maximizes the spectral efficiency is spatial multiplexing and, after 1998, the atmosphere was therefore primed for it to finally come to the fore.

The prerequisite for spatial multiplexing is a multiple-input multiple-output (MIMO) communication channel, where each input/output refers to an antenna element in a digital antenna array.
There are two MIMO categories: single-user MIMO entails a multiantenna BS and a multiantenna user device, while multiuser MIMO encompasses a multiantenna BS and multiple user devices.

Arguably, the main catalyst for single-user MIMO was the work in \cite{Foschini1998a}, which set out to design the perfect transceivers from an information-theoretic standpoint. Starting with transmit and receive digital antenna arrays and no preset conditions on how to employ them, it was found that, if the elements within each array exhibited uncorrelated fading, the optimum strategy was to have each radiate an independent data-carrying signal.
This was radically novel in that it sought to exploit, rather than counter, multipath propagation; it is the very existence of multiple paths that allows the receiver to observe a distinct linear combination of the transmit signals at each receive antenna, where-from those transmit signals can be resolved. The number of signals that can be spatially multiplexed is then limited by the minimum of the number of transmit and receive antenna elements.
In follow-up work, a specific architecture was proposed to effect such spatial multiplexing, the so-called layered architecture, which was remarkable in that it could be built with off-the-shelf encoders and decoders and did not require the transmitter to know anything about the channel  \cite{WolFosGol:V-BLAST:-an-architecture-for-realizing:98}.
Additional results progressively solidified the theoretical underpinnings \cite{Telatar1999a}.
In particular, the idea of transmitting concurrent signals, one from each antenna element, was generalized to the transmission of concurrent beams from all elements at once. Phased arrays cannot achieve such spatial multiplexing because they only create one beam at a time, and digital antenna arrays are decidedly necessary.

Multiuser MIMO can be traced back to signal processing concepts for simultaneous uplink reception from multiple users \cite{Winters1987a} and simultaneous downlink beamforming to users in different angular directions \cite{Swales1990a}.
Here, the number of signals that can be spatially multiplexed is not limited by the number of antenna elements per user, but rather across all users; even if each user features a single element, it is possible to spatially multiplex one signal to/from each one. This major advantage comes at the expense of the BS having to carefully arrange the transmit and receive beams such that each one matches with the multipath characteristics of its intended user and there is minimal interference among them, as in Fig.~\ref{fig:array_antennas}(b).
With that, every user can transmit continuously and over the entire system bandwidth, rather than only in a time slot and/or frequency subband, reflecting the spatial multiplexing benefit.
Multiuser MIMO is a generalization to multipath settings of classical space-division multiple access (SDMA), whereby users share a channel in space rather than in time or frequency. Interestingly, the SDMA concept is more than 20 years older than single-user MIMO \cite{Tsuji1974a}, which showcases that  establishing many simultaneous user connections was long perceived more important in wireless networks than achieving high data rate per connection.

The potential of MIMO, in both its single-user and multiuser fashions, sparked a chain reaction that spread rapidly through academia and industry, bringing much excitement by the early 2000s.
Cellular standardization bodies, in particular, the 3G partnership project (3GPP), adopted it in a limited fashion for late 3G releases and then as an integral part of the designs beginning with 4G.
Even faster was the adoption within Wi-Fi, with the first version including MIMO certified in 2007 and supported by a multitude of devices including laptops, tablets, and smartphones.

\begin{figure}[t!]
        \centering 
	\begin{overpic}[width=.7\columnwidth,tics=10]{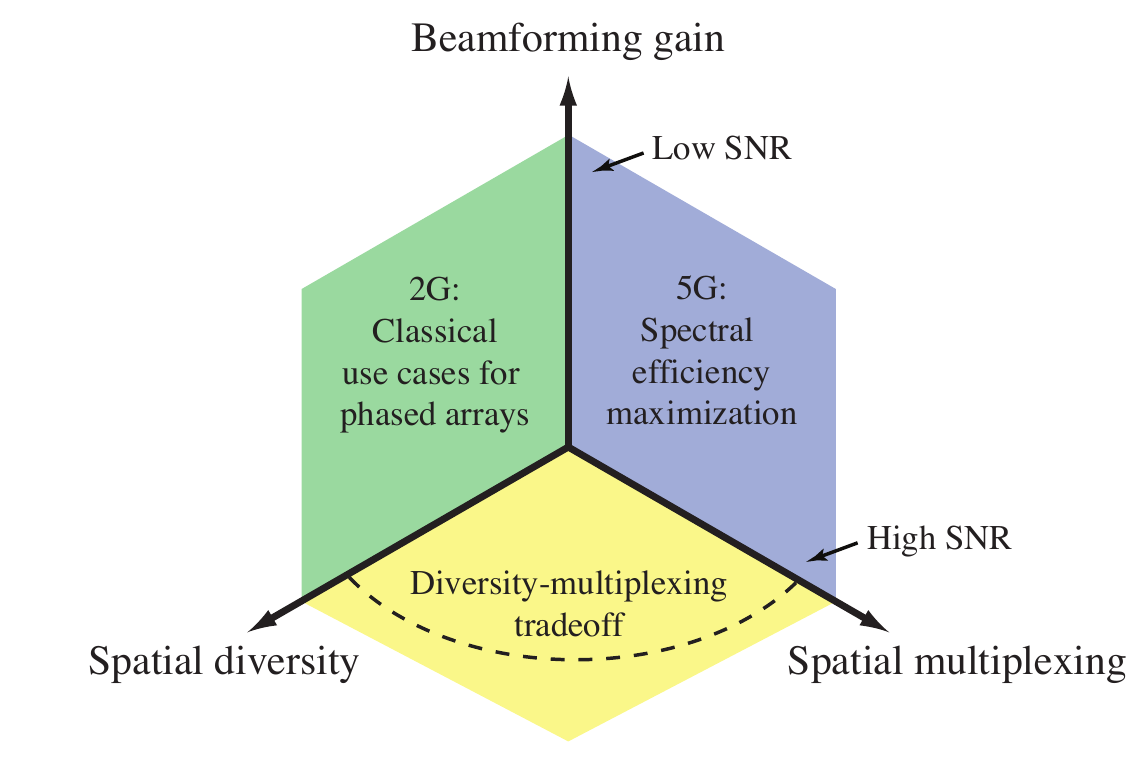}
\end{overpic} \vspace{-5mm}
        \caption{There are three benefit dimensions of multiantenna communication that have developed in past decades. From 2G to 5G, systems have shifted from the green plane to the blue plane; that is, spatial diversity has gradually been replaced by spatial multiplexing. Since spatial multiplexing requires high SNRs to be practically useful, beamforming gains remain essential at low SNRs.} 
        \label{fig:three_benefits}  
\end{figure}

MIMO harnesses the three dimensions of benefits shown in Fig.~\ref{fig:three_benefits}:
beamforming gain, spatial diversity, and spatial multiplexing. A clear understanding of how these benefits are related has emerged over time. Fundamental trade-offs have been identified, for given array configurations and channel conditions.
\begin{itemize}
    \item Beamforming is a special case of spatial multiplexing where a single beam is transmitted to a single user. This is in fact the optimum strategy when the signal-to-noise ratio (SNR) is low; maximizing the signal energy is then of the essence, and the best recipe is to concentrate all the radiated energy on the strongest beam. At high SNR, in contrast, energy is plentiful and can be spread over multiple beams, to the point that it is optimum to activate as many beams as the channel and antenna counts allow. This is represented by the blue plane in Fig.~\ref{fig:three_benefits}.
    \item Spatial diversity, roughly quantified as the number of independently faded signal replicas, and spatial multiplexing, meaning the number of concurrent beams, cannot be simultaneously maximized. These two quantities are rather subject to a tradeoff \cite{Zheng2003a}. At the extreme points of this diversity-multiplexing tradeoff, one of the quantities is maximized while the other stands at a minimum. Various combinations are feasible at intermediate points of the tradeoff curve, which is cartooned on the yellow plane of Fig.~\ref{fig:three_benefits}.
\end{itemize}
As mentioned, the choice between beamforming and spatial multiplexing is dictated by the SNR, hence, by its underlying parameters (e.g., transmit power, channel attenuation, noise power). In turn, the mix of diversity and multiplexing depends on whether the priority is to increase reliability or spectral efficiency. However, the operating point should be selected holistically, and over the years this has caused the mix to shift towards less diversity and more multiplexing. Indeed, as successive system generations have spanned ever broader bandwidths, more and more diversity has been reaped in the frequency domain.
The rewards of additional diversity rapidly saturate, thus the need for spatial diversity has abated \cite{Lozano2010a}. This of course does not apply to narrowband control channels or to low-power short-packet IoT communication, where spatial diversity remains important, but it does hold for the user data channels that carry the bulk of the traffic in cellular networks.

\subsection*{From Spatial Multiplexing to Massive MIMO}

The basics of MIMO, developed under the premises of perfect channel state information (CSI) and rich scattering, indicate that an arbitrarily high spectral efficiency can be achieved by deploying sufficiently many antennas and serving many users at once.
However, the practical challenges became apparent when the technology was first commercialized.
The spatial multiplexing capability in single-user MIMO was often restricted by limited scattering, while multiuser MIMO is restrained by imperfect CSI.
Massive MIMO, a new form of multiuser MIMO that originated from  \cite{Marzetta2010a}, was developed in the 2010s to address these issues and is now at the heart of 5G. 

The new aspects of massive MIMO are as follows.
First, it relies on having many more BS antennas than spatially multiplexed users. This design choice renders the beams relatively narrow (e.g., in the sense of focusing on a small region around the intended receiver), hence there is likely to be little overlap among beams focused on distinct users.
Moreover, by virtue of these \emph{favorable} conditions, whatever little interference exists can be suppressed through
low-complexity linear signal processing; for example, regularized zero-forcing  that fine-tunes each beam's focal area to balance a strong beamforming gain with low interference \cite{Joham2005a}.
At the same time, and again because of the excess BS antennas, the effective channels provided by these beams \emph{harden}, meaning that they become very stable and subject to only minimal fading fluctuations.

Second, massive MIMO is tailored for resource-efficient CSI acquisition. The main estimation principle is to emit separate predefined pilot signals from each antenna element and then gauge the channel coefficients from the observations of these pilots at the receive elements.
Massive MIMO adopted time-division duplexing (TDD), where the same bandwidth is utilized, in alternating fashion, for uplink and downlink.
Since, by virtue of reciprocity, the channel is then identical in both directions, it suffices to estimate its coefficients in one direction. Specifically, the CSI required for both uplink and downlink is obtained from uplink pilots.
The necessary pilot resources are thus determined by the number of multiplexed users, with no dependence on the number of BS antennas. 
In contrast, many previous commercial implementations of multiuser MIMO were based on frequency-division duplexing (FDD), where the uplink and downlink channels were entirely different, or TDD  operation without using reciprocity. The downlink operation then required the BS to transmit as many pilots as it has antennas, except in specific propagation scenarios where the channels can be parametrized using a few angles. Moreover, each user needed to quantize and feed back its channel estimates to the BS. In a typical 5G setup with $M=64$ BS antennas that spatially multiplex $K=8$ users, the FDD alternative would require $M/K=8$ times as many pilots and a proportional amount of extra CSI feedback.

\begin{figure}[t!]
        \centering 
	\begin{overpic}[width=.6\columnwidth,tics=10]{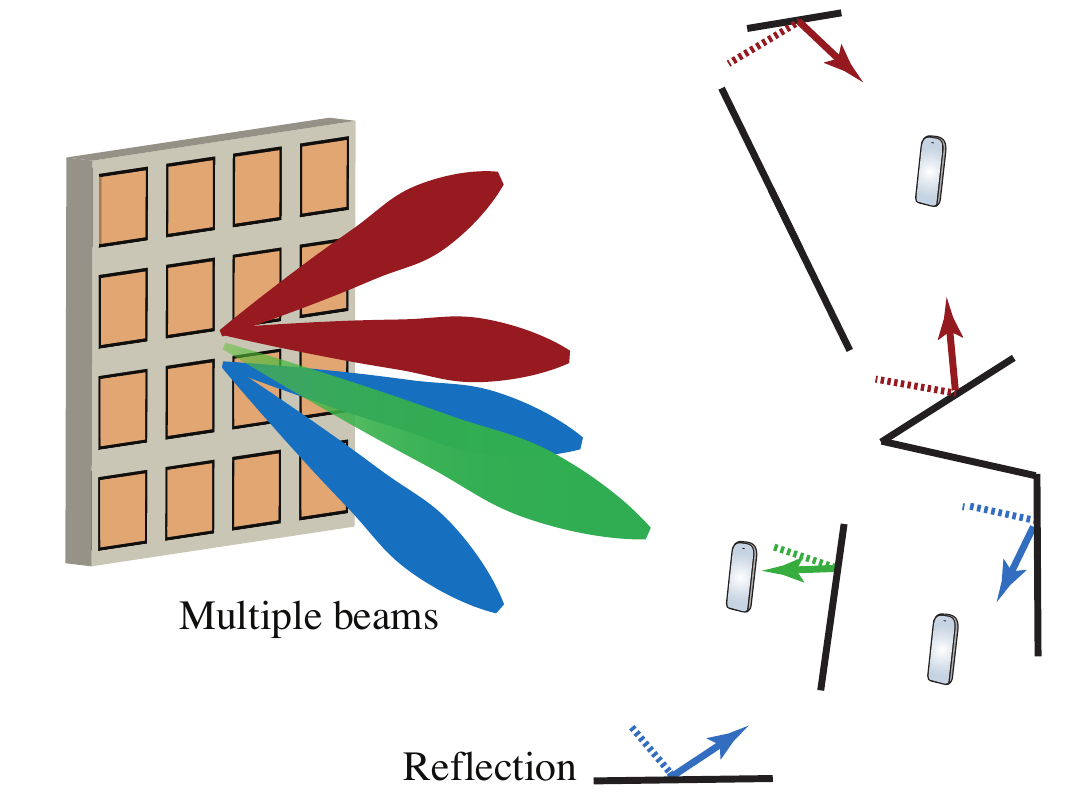}
\end{overpic} \vspace{-3mm}
        \caption{The propagation channel consists of a multitude of specularly or diffusely reflecting objects. Such channels are very challenging to estimate with sufficient accuracy to enable multiuser MIMO communication, but the TDD CSI estimation approach in massive MIMO manages this.} 
        \label{fig:beamforming_multipath}  
\end{figure}

The TDD operation is particularly helpful in complex propagation environments with many paths per user, such as the one sketched in Fig.~\ref{fig:beamforming_multipath}, where the optimum downlink transmission spreads a user's signal energy in many directions to match the reflecting objects.
CSI acquisition through uplink pilots automatically captures these fine characteristics, without any prior channel knowledge or array calibration.
In FDD operation, besides requiring vastly more pilot resources, essential channel details are
lost in the feedback quantization.
In cellular networks, pilot signals must be reused with care across cells to avoid \emph{pilot contamination} phenomena whereby BSs inadvertently beamform towards pilot-sharing users in neighboring cells.
This is particularly a concern in TDD operation, where uplink estimation errors also affect the downlink. A multitude of signal processing and resource allocation schemes  have been developed over the past decade to alleviate pilot contamination \cite{Marzetta2016a,massivemimobook}.

Massive MIMO provides a solid foundation for practical signal processing design.
While sophisticated information theory for multiuser MIMO existed already by the 2000s \cite{weingarten2006capacity}, it was largely limited to scenarios with perfect CSI. As CSI quality is the main limiting factor of multiuser MIMO performance, this constrained the practical usefulness of the available theory. Thanks to the reliance on linear signal processing, massive MIMO analyses successfully handle imperfect CSI and hardware imperfections, resulting in rigorous and mathematically clean spectral efficiency expressions that not only predict actual performance accurately, but serve as effective tools for system optimization (e.g., pilot allocation, power control, and beamforming). Massive MIMO theory not only turned multiuser MIMO into a practically feasible technology; the analytical elegance also expanded the way information theory for wireless communication can be taught \cite{Marzetta2016a,massivemimobook}.

As mentioned, uplink-downlink reciprocity in TDD operation is important for massive MIMO. Reciprocity holds for the over-the-air propagation as long as the channel impulse response remains constant; that is, provided the duplexing takes place within the channel coherence time. However, the transceiver hardware is generally not reciprocal between transmission and reception, for instance due to mismatches in the local oscillators. 
Such hardware non-reciprocity calls for a calibration procedure that phase-synchronizes the antennas within each array through occasional mutual measurements.

Today, 5G BSs feature almost exclusively massive MIMO configurations in TDD bands, with arrays of $M=32$ or $M=64$ antennas being the most common. Early on, there were concerns that the signal processing would entail an exceedingly high energy consumption, but this concern was later dispelled, and dedicated systems-on-chip are now available that implement clever signal processing algorithms for the entire BBU, including massive MIMO, at reasonable energy costs.

\subsection*{A Quest for More Bandwidth at Higher Frequencies}

The bit rate has long been the performance metric that users of wireless technology are most familiar with,
and hence wireless technology has evolved
to support higher values thereof.
The bit rate enjoyed by a single device equals the product of the spectral efficiency and the spectral bandwidth. Therefore,
besides being driven higher by single-user MIMO, bit rates have expanded over time thanks to the allocation of new frequency bands.
A 2G network typically had access to $20$ MHz at carrier frequencies around $1$ GHz, while current 5G networks primarily span $100$ MHz in the $3.5$-GHz range, with the standard supporting in excess of $500$ MHz.

The radio spectrum is a limited natural resource shared by a multitude of technologies, including those beyond the civilian wireless communication arena considered in this article.
While a few sub-$6$ GHz bands have been refarmed from outdated technologies to cellular networks, the strive for fresh bandwidth inevitably pushes systems towards ever higher frequencies.
In particular, millimeter-wave (mmWave) bands, nominally starting at $30$ GHz, are now part of 5G.
First-generation mmWave technology has been rolled out by a few telecom operators, while the bulk of them wait for the hardware to mature and for the $3.5$-GHz band to become congested.

The field strength of a signal radiated from a point source in free space attenuates with distance in a frequency-independent manner. 
Then, the power captured from such an electric field is proportional to the receiver's aperture and, 
since the size of an antenna element shrinks with the wavelength, an increased carrier frequency necessitates further antenna elements to maintain the desired aperture.
Moreover, the channel conditions in cellular networks become steadily more challenging as the frequency shifts up due to reduced scattering and diffraction, and steeper penetration losses, all of which call for beamforming gains.
Multiantenna technology is therefore paramount at mmWave frequencies.

At the same time, implementation becomes difficult, and not only because of the added hardware components and the huge dimensionalities in digital signal processing.
When moving to higher frequencies and broader bandwidths, hence to faster sampling rates, power amplifier efficiency and dissipation in analog-to-digital converters (ADCs) are further issues that need attention.
The signal processing community has explored two main ways to deal with the hardware and algorithmic complexity~\cite{heath2016overview}.

\begin{figure}[t!]
        \centering 
	\begin{overpic}[width=.95\columnwidth,tics=10]{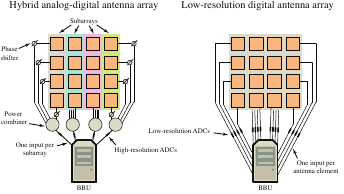}
\end{overpic} 
        \caption{Conventional digital antenna arrays incur a high energy consumption when implemented at mmWave frequencies. There are two main ways to circumvent this: 1) Reduce the number of components using hybrid analog-digital antenna arrays, consisting of multiple phased subarrays; 2) Design digital antenna arrays with simplified components, such as low-resolution ADCs.} 
        \label{fig:hybrid_array_antenna}  
\end{figure}

The first option is to reduce the number of RF units, particularly converters, by designing transceivers as a mix of phased arrays and digital antenna arrays. The resulting hybrid analog-digital antenna array is illustrated in Fig.~\ref{fig:hybrid_array_antenna}, where each column is a phased subarray that is connected separately to the BBU, such that different signals can be transmitted and received. 
Each subarray can form a single beam and spatial multiplexing can then be applied through different linear combinations of those beams. 
If the channel features a small number of propagation paths, each subarray can focus a beam on one of those paths, and the communication performance of a digital antenna array can be attained with fewer hardware components.
In the multipath scenario illustrated in Fig.~\ref{fig:beamforming_multipath}, five distinct beam directions are sufficient to communicate effectively.
Hybrid antenna arrays can take other forms, but generally entail a semi-analog beamforming implementation with more antenna elements than digital ports \cite{mendez2016hybrid}.
There are several prices to pay for abandoning the digital antenna array paradigm.
One can only transmit as many beams as there are digital ports and the beamforming fidelity is crippled in wideband systems since combinations of the same beams must be used on all subcarriers. Channel estimation becomes more intricate since each phased array must sweep through as many beam directions as it has elements in order to excite all channel dimensions.

An alternative to reducing the number of RF units is digital antenna arrays with lowered ADC resolutions, as also illustrated in Fig.~\ref{fig:hybrid_array_antenna}. The energy consumption of an ADC grows exponentially with the resolution, hence enormous energy reductions are possible by moving from the conventional $15$ bit/sample down to, say, $5$ bits/sample. And yet, since an array with $M$ elements and $b$-bit ADCs collects a total of $bM$  bits per sample period, the total number of ADC bits can still be sizeable even if $b$ is small, explaining why a high spectral efficiency can be maintained \cite{Mollen2017a}.
The extreme case of uplink massive MIMO with $b=1$  happens to be analytically tractable \cite{Li2017a}, which has facilitated the emergence of signal processing algorithms that compensate for the ensuing quantization distortion.
The downlink counterpart involves low-resolution
digital-to-analog converters (DACs) \cite{Jacobsson2019a}.

Reduced bit resolution is also viable for hybrid antenna arrays; if the analog signal combining prior to quantization and the digital post-processing is properly optimized for the communication task at hand, the signal content becomes more amendable to a low-bit representation \cite{SE20}.

The initial 5G mmWave products are based on hybrid arrays, but there are indications that the low-resolution approach might eventually become the preferred solution \cite{roth2018comparison}.

\subsection*{Further Opportunities for Dimensionality Reduction}

The joint evolution towards arrays with more antenna elements and wider bandwidths, requiring higher sampling rates, makes it essential not to overdesign the transceivers. As an alternative to scaling up a conventionally small digital array, Fig.~\ref{fig:hybrid_array_antenna} showcased two ways of increasing the antenna element counts while reducing the hardware complexity per element.
Both approaches capitalize on the massive MIMO philosophy of having more antenna elements than multiplexed beams, which enables a reduction in the beamforming exactness because the beams are so narrow that inter-user interference is anyway low.
Further dimensionality reductions are possible by exploiting the structure of the channel, say sparsity in the angular, frequency, and time domains, or by exploiting the specific task the system is designed to address.

\begin{figure}[t!]
        \centering 
	\begin{overpic}[width=.95\columnwidth,tics=10]{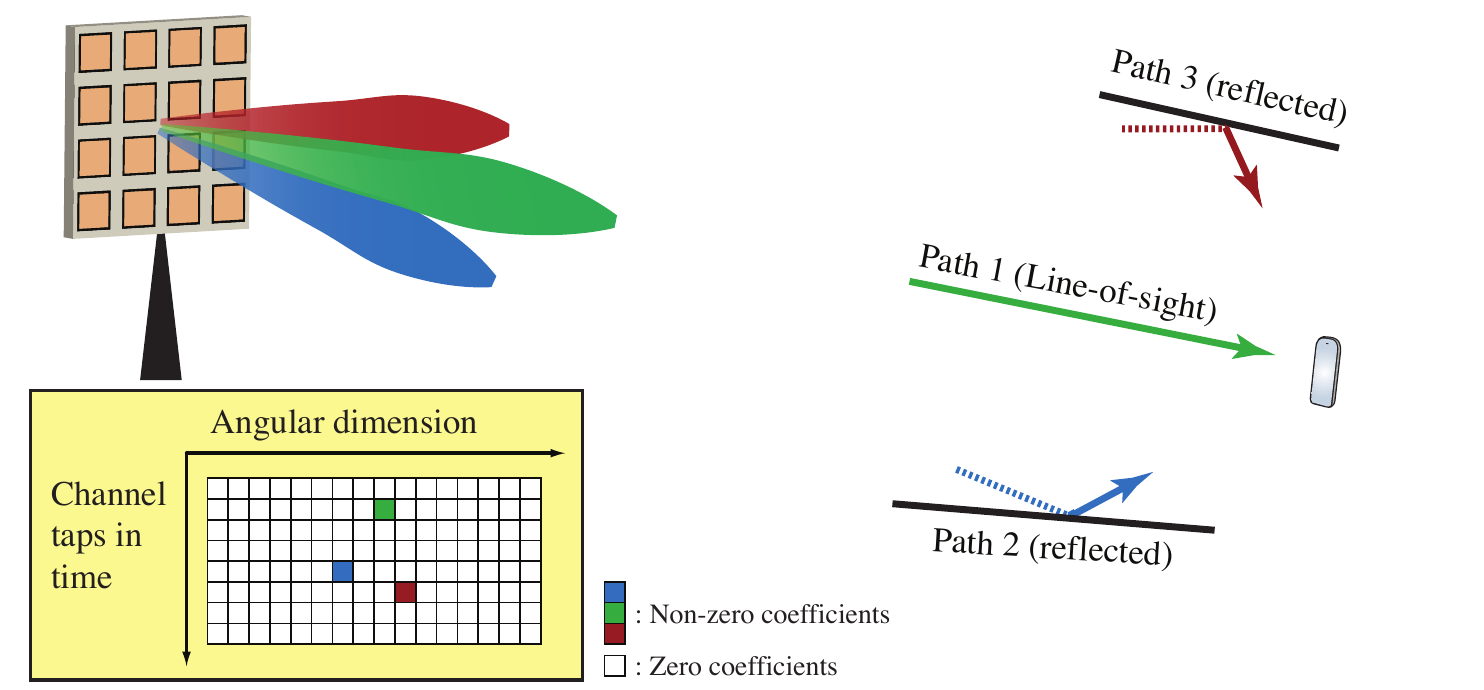}
\end{overpic} 
        \caption{The channel in mmWave systems with large spectral bandwidths and antenna counts might exhibit sparsity in the joint time-angle domain. In this example, there are $N=3$ paths that are distinct in both time and angle. The sparse impulse response can be exploited along with compressed sensing techniques to reduce the sampling rate, thereby lowering power consumption.} 
        \label{fig:beamforming_sparsity}  
\end{figure}

Besides exponential in the resolution, the energy consumption of an
ADC is proportional to the sampling rate. A host of ideas based on sub-Nyquist sampling and compressed sensing have been proposed to reduce the sampling rate by exploiting various forms of channel structure that exist in many scenarios \cite{eldar2015sampling}.
In mmWave channels with a small number $N$ of propagation paths, there might be only roughly $N$ nonzero taps in the channel impulse response regardless of the bandwidth. 
Such time-domain sparsity in the channel response can be leveraged to reduce the sampling rate \cite{GDCS18}.
The $N$ paths are likely distinct also in the angular domain, given that BSs are usually deployed high above the environmental clutter; reflections take place only locally around each user, subtending a small angular spread at the distant BS.
Fig.~\ref{fig:beamforming_sparsity} illustrates such a scenario with $N=3$ distinct paths, supporting three multiplexed signals. The line-of-sight path has a distinctly short delay, while the two remaining paths have similar delays but are clearly distinguishable in the angular realm. The joint channel sparsity is represented by the three colored entries in a time-angle matrix, with the vast majority of entries
containing no propagation paths. 
Combining these forms of sparsity with modern compressed-sensing tools enables hefty reductions in sampling rates.

Many data services exhibit intermittent activity patterns; among the thousands of devices associated with a BS, only a small subset requires data transfers within a given time slot. Signal processing can enable these devices to transmit efficiently without requiring a preceding access procedure.
The key is to assign each device with a unique but non-orthogonal pilot sequence and then utilize sparsity in the user domain along and the large number of spatial samples obtained over an antenna array to enable user identification and channel estimation \cite{Liu2018a}.
The joint user and data detection problem has also been approached using compressed sensing methods \cite{Yuan2020a}.

Commercial massive MIMO products already exploit some elementary channel sparsity; for instance, the received signals over the many antennas might be transformed into an equal number of angular dimensions. The dimensions that contain little power are discarded in the early stages of the digital uplink processing to shrink the dimensionality of the remaining computations. However, the more radical compressed sensing solutions are yet to be brought to life.

\subsection*{Machine-Learning-Based Algorithmic Refinements}

One of the most active areas in contemporary signal processing is machine learning (ML). While this is a many-decades-old discipline, the increased availability of large amounts of data and processing power has, in recent years, greatly enhanced its potential to transform the implementation of many signal-processing tasks from more traditional model-driven algorithms into data-driven ones.  This transformation is also taking place in the context of signal processing for wireless communications, which has traditionally been very heavily (and successfully) model-based. Several trends are driving this transformation. One trend is that, with the vast amount of IoT and machine-type connections that coexist with human-type broadband connections, wireless network traffic is becoming increasingly intricate to model accurately, thereby making network operation difficult to optimize. Another trend is that antenna arrays and other sensors are becoming pervasive on smartphones and other connected devices, hence the volume of data available for learning is swelling dramatically. Yet a third trend is that the amount of processing power distributed throughout wireless networks is growing rapidly, giving rise to paradigms such as fog and edge computing.

There is a confluence of ML and communications in the optimization of wireless networks. This is a very natural application for ML since the operation of these networks involves a multitude of tasks that ML is good at addressing, including inferential tasks such as channel estimation, signal detection, and data decoding, as well as decision-making tasks such as routing, access control, and resource allocation. 
ML-based solutions can capture practical characteristics that were overlooked by the models underpinning existing algorithms.
However, to ensure that ML algorithms improve upon the existing, it is essential to initiate the training procedure judiciously.

\begin{figure}[t!]
        \centering 
	\begin{overpic}[width=\columnwidth,tics=10]{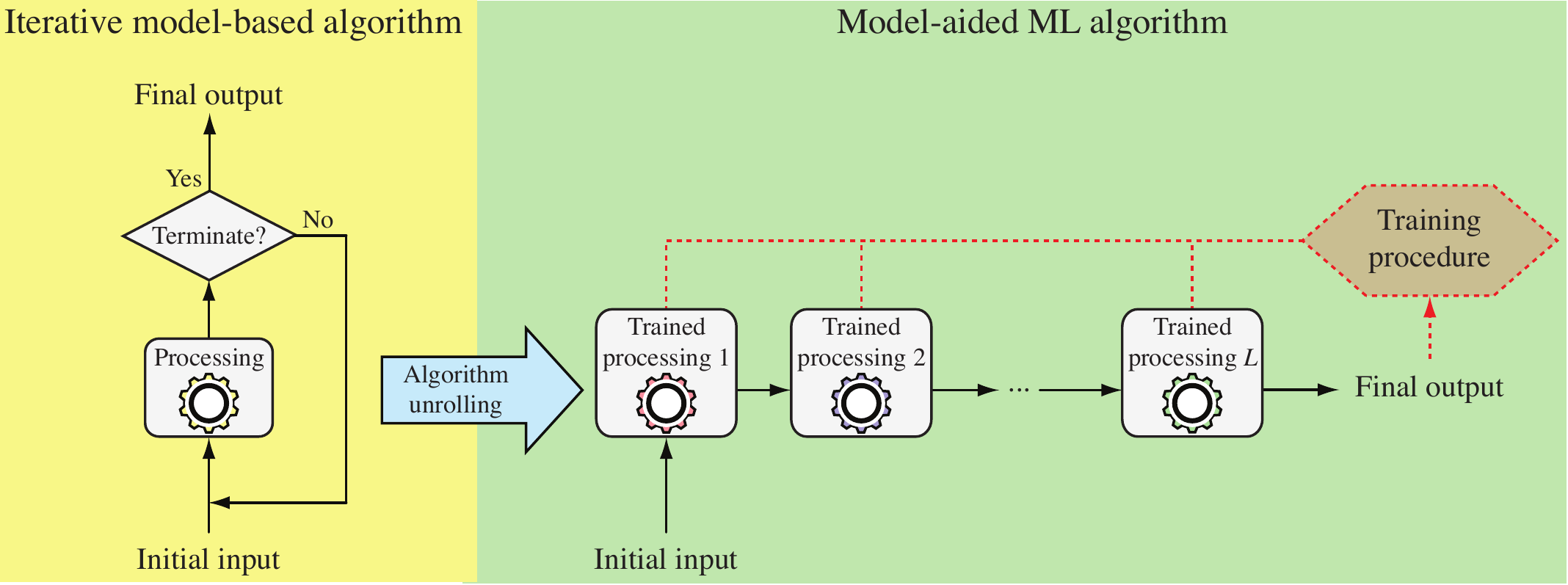}
\end{overpic} 
        \caption{Many conventional model-based algorithms for optimization in multiantenna communication are iterative. Suppose one such algorithm can be expressed as an iterative processing loop that continues until a termination criterion is satisfied.
        An enhanced ML algorithm can be developed through algorithm unrolling; that is, writing the iteration as $L$ separate processing layers and fine-tuning these layers through a data-aided training procedure.} 
        \label{fig:unrolling}  
\end{figure}

The \emph{model-aided ML paradigm} provides a structured way to transfer classical know-how from the signal processing community onto new ML algorithms \cite{Monga2021a}. Fig.~\ref{fig:unrolling} exemplifies how an existing iterative algorithm can be transformed into an enhanced ML algorithm.
The existing algorithm takes an initial input signal and processes/updates it iteratively until a predefined termination criterion is satisfied, at which point the final output is obtained. 
The specific processing is normally obtained through model-based algorithm design. Instead of expressing the algorithm as a loop,
$L$ iterations of the algorithm can be expressed as a sequence of $L$ identical processing layers.
If a data-driven training procedure is employed to fine-tune
these processing layers, which no longer have to be identical, what ensues is an ML algorithm that is guaranteed to perform better than the original model-based algorithm. This procedure is called \emph{algorithm unrolling} or deep unfolding, and it has in recent years been utilized to enhance various multiantenna tasks, including signal detection \cite{Samuel2019a}
and beamforming optimization for downlink multiuser MIMO \cite{Pellaco2022a}.

\section*{A Peek into the Future}

Over the past 25 years, multiantenna techniques have gone from rudimentary designs for beamforming and diversity combining to a mainstream technology that uses massive spatial multiplexing to multiply the capacity of 5G networks. The MIMO technology is now present in every smartphone and BS that enters the market. 
Fast and capable signal processing algorithms have enabled this leap forward, and are currently buttressing the emergence of low-power 5G mmWave transceivers where high-resolution hardware components are replaced with digital processing. Recent theoretical breakthroughs, including ML-based algorithms, are bound to continue sustaining the progress of the technology.

Indeed, we expect the multiantenna communication journey to continue. The insatiable growth in data traffic can only be met by deploying ever more antennas and ever more bandwidth.
The massive MIMO philosophy prescribes that the number of BS antennas, $M$, must scale proportionally to the number of active users, $K$. As the complexity of algorithms such as regularized zero-forcing is proportional to $M K^2$ \cite{massivemimobook}, a linear scaling in both $K$ and $M$ implies a cubic complexity growth. Moreover, once the bandwidth surpasses $1$ GHz, the sampling rates approach the clock speed of existing processors, which renders the implementation even more demanding.
The compressed sensing algorithms described earlier might be suitable to address these challenges, but there is likely room for many new signal processing advances.

When adding ever more antennas to BSs, practical size and weight constraints might make new deployment principles necessary, beyond the boxes-in-a-tower paradigm. One promising approach is to distribute the antennas over multiple physical locations while retaining the coherent transmission and reception processing, a concept rooted in cell cooperation and network MIMO ideas, as well as in the notion of remote RF units, and whose present embodiment is termed cell-free massive MIMO \cite{cellfreebook}.
Apart from stronger beamforming gains, a distributed antenna deployment can provide improved spatial multiplexing capabilities and macroscopic diversity against the shadowing of large objects in the environment. The current trend of shifting baseband computations from BS sites to edge-cloud computers will ease the adoption of this deployment approach.

After two decades of smartphones ruling the wireless ecosystem, other devices such as extended reality eyeglasses are predicted to
take center stage.
New services will surface, with renewed standards for the bit rates, latency, and reliability that users expect wireless networks to deliver. Other performance metrics might arise to dictate future technology development, particularly related to sustainability, environmental impact, and deployment costs, as well as to the digital divide between the digitized and far-from-digitized regions of the world.

Beyond the signal processing advances captured in this article, two emerging research topics build on multiantenna technology. The first is integrated communication and sensing \cite{Zhang2021a}, which explores how large-scale antenna arrays can be simultaneously used for accurate radar sensing, localization, and communication. It seems natural that the deployment of massive antenna numbers for communication purposes can be the catalyst for other applications that benefit from wireless measurements.
Another related research direction is that of smart surfaces \cite{Bjornson2022a}, whose signal reflection properties can be controlled by means of metamaterials with programmable impedance patterns.
These reconfigurable intelligent surfaces provide a sort of passive beamforming that is particularly useful to enhance propagation conditions over wireless channels.

\section*{Authors}

\textbf{Emil Björnson} (emilbjo@kth.se) is a Professor of Wireless Communication at the KTH Royal Institute of Technology, Stockholm, Sweden. He has authored three textbooks on MIMO technology, has received 23000 citations, and has published a large amount of simulation code. He has received the 2018 and 2022 IEEE Marconi Prize Paper Awards in Wireless Communications, the 2019 EURASIP Early Career Award, the 2019 IEEE Communications Society Fred W. Ellersick Prize, the 2019 IEEE Signal Processing Magazine Best Column Award, the 2020 Pierre-Simon Laplace Early Career Technical Achievement Award, the 2020 CTTC Early Achievement Award, and the 2021 IEEE ComSoc RCC Early Achievement Award. He is an IEEE Fellow.

\textbf{Yonina C. Eldar} (yonina.eldar@weizmann.ac.il) is a Professor in the Department of Math and Computer Science at the Weizmann Institute of Science, Rehovot, Israel, where she heads the Center for Biomedical Engineering and Signal Processing. She is also a Visiting Professor at MIT and at the Broad Institute and an Adjunct Professor at Duke University, and was a Visiting Professor at Stanford University. She is a member of the Israel Academy of Sciences and Humanities and a EURASIP Fellow. She has received many awards for excellence in research and teaching, including the IEEE Signal Processing Society Technical Achievement Award, the IEEE/AESS Fred Nathanson Memorial Radar Award, and the IEEE Kiyo Tomiyasu Award. She heads the Committee for Promoting Gender Fairness in Higher Education Institutions in Israel. She is an IEEE Fellow. 

\textbf{Erik G. Larsson} (erik.g.larsson@liu.se) is Professor at Link\"oping University, Sweden.  He co-authored the textbook \emph{Fundamentals of
  Massive MIMO} (Cambridge University Press, 2016). He received,
among others, the IEEE ComSoc Stephen O. Rice Prize in Communications
Theory in 2015, the IEEE ComSoc Leonard G. Abraham Prize in 2017, the
IEEE ComSoc Best Tutorial Paper Award in 2018, and the IEEE ComSoc
Fred W. Ellersick Prize in 2019.  His interests include wireless
communications, statistical signal processing, and networks. He is an IEEE Fellow.

\textbf{Angel Lozano} (angel.lozano@upf.edu) is a Professor at Universitat Pompeu Fabra. He received his Ph.D. from Stanford University in 1999, worked for Bell Labs (Lucent Technologies, now Nokia) between 1999 and 2008, and served as Adj. Associate Professor at Columbia University between 2005 and 2008.  His papers have received several awards, including the 2009 Stephen O. Rice prize, the 2016 Fred W. Ellersick prize, and the 2016 Communications Society \& Information Theory Society joint paper award. He is also the recipient of an ERC Advanced Grant for the period 2016-2021 and a 2017 Highly Cited Author. He is the coauthor of the textbook “Foundations of MIMO Communication,” published by Cambridge University Press in 2019. He is an IEEE Fellow.

\textbf{H. Vincent Poor} (poor@princeton.edu) is the Michael Henry Strater University Professor at Princeton University, where his interests include information theory, machine learning, and network science, and their applications in wireless networks, energy systems, and related fields. Among his publications in these areas is the recent book \emph{Machine Learning and Wireless Communications} (Cambridge University Press, 2022). He is a member of the U.S. National Academies of Engineering and Sciences and received the IEEE Alexander Graham Bell Medal in 2017. He is an IEEE Life Fellow.

\bibliographystyle{IEEEtran}
\bibliography{IEEEabrv,refs}

\end{document}